# SUSY Dark Matter in Universal and Nonuniversal Gaugino Mass Models

D. P. Roy

Homi Bhabha Centre for Science Education,

Tata Institute of Fundamental Research, Mumbai 400088, India

**Abstract**

We review the phenomenology of SUSY dark matter in various versions of MSSM, with universal and nonuniversal gaugino masses at the GUT scale. We start with the universal case (CMSSM), where the cosmologically compatible dark matter relic density is achieved only over some narrow regions of parameter space, involving some fine-tuning. Moreover, most of these regions are seriously challenged by the constraints from collider and direct dark matter detection experiments. Then we consider some simple and predictive nonuniversal gaugino mass models, based on SU(5) GUT. Several of these models offer viable SUSY dark matter candidates, which are compatible with the cosmic dark matter relic density and the above mentioned experimental constraints. They can be probed at the present and future collider and dark matter search experiments. Finally, we consider the nonuniversal gaugino mass model arising from anomaly mediated SUSY breaking. In this case the cosmologically compatible dark matter relic density requires dark matter mass of a few TeV, which puts it beyond the scope of collider and direct dark matter detection experiments. However, it has interesting predictions for some indirect dark matter detection experiments.

## 1. Introduction:

The most phenomenologically attractive feature of Supersymmetry (SUSY) [1] is that it provides a natural candidate for the dark matter of the universe [2] in terms of the lightest supersymmetric particle (LSP). We shall consider here the most popular version of SUSY, called the minimal supersymmetric standard model (MSSM), which consists of the standard model particles along with their supersymmetric partners. The astrophysical constraints on the dark matter requires it to be a colourless and chargeless particle, while direct dark matter detection experiments disfavour it to be the supersymmetric partner of neutrino, called sneutrino $\tilde{\nu}$. Thus the leading candidate for the LSP dark matter in this model is the lightest neutralino $\tilde{\chi}_1^0$, abbreviated as χ. It is the supersymmetric partner of one of the electroweak gauge bosons B, W, or a Higgs boson H, called bino, wino or higgsino respectively. In general it can be an admixture of these states, i.e.

$$\chi \equiv \tilde{\chi}_1^0 = c_1 \tilde{B} + c_2 \tilde{W} + c_3 \tilde{H}_d + c_4 \tilde{H}_u \quad , \tag{1}$$

where the subscripts u and d refer to the two Higgs doublets of MSSM, giving mass to the up and down type fermions.

The 4x4 neutralino mass matrix in this basis is

$$M_N = \begin{pmatrix} M_1 & 0 & -M_Z \sin\theta_W \cos\beta & M_Z \sin\theta_W \sin\beta \\ 0 & M_2 & M_Z \cos\theta_W \cos\beta & -M_Z \cos\theta_W \sin\beta \\ -M_Z \sin\theta_W \cos\beta & M_Z \cos\theta_W \cos\beta & 0 & -\mu \\ M_Z \sin\theta_W \sin\beta & -M_Z \cos\theta_W \sin\beta & -\mu & 0 \end{pmatrix}, \tag{2}$$

where $\theta_W$ is the weak mixing angle and tanβ represents the ratio of the two Higgs vacuum expectation values. The diagonal elements are $M_1$, $M_2$ and $\pm \mu$ in the mass basis of $\tilde{B}, \tilde{W}$ and $\tilde{H}_{1,2} = \tilde{H}_d \pm \tilde{H}_u$. Note that all the off-diagonal elements are less than the Z boson mass $M_Z$. There are experimental indications that the SUSY masses represented by these diagonal elements are at least $> 2M_Z$ [3]. This means that the mixing angles are small, so that the LSP χ is dominated by one of these interaction eigenstates.

There is one exception to this, i.e.

$$M_{ii} \approx M_{jj} \Rightarrow \tan 2\theta_{ij} = 2M_{ij}/(M_{ii} - M_{jj}) : l \arg e \Rightarrow \chi = \tilde{B} - \tilde{H}, \tilde{W} - \tilde{H}. \tag{3}$$

When two diagonal elements are large but nearly degenerate, then the corresponding mixing angle can be large; and thus the LSP dark matter χ can be a mixed bino-higgsino or wino-higgsino state. This has been referred to as the well-tempered neutralino scenario [4]. We shall see below that an important example of this is realized in the focus point region of the universal gaugino mass model [5]. But more generally one expects the LSP dark matter χ to approximately correspond to one of the interaction eigenstates.

We shall see below that the cosmologically compatible dark matter relic density has strong constraints on the underlying SUSY model. The most robust of these constraints is an upper limit of a few TeV on the LSP (χ) mass, which implies observable SUSY signals in the future collider and/or dark matter experiments. This is particularly compelling in the case of the universal gaugino mass model, which is also called the constrained MSSM (CMSSM) or the minimal supergravity (mSUGRA) model. This will be the subject of our investigation in the next section.

## 2. Universal Gaugino Mass Model (CMSSM or mSUGRA):

In this model the SUSY breaking in the hidden sector is communicated to the observable sector via gravitational interaction at the high energy scale of grand unified theory (GUT), where the SU(3)xSU(2)xU(1) gauge couplings $\alpha_{3,2,1}$ of the standard model are unified to common value $\alpha_G$. Since the gravitational interaction is colour and flavour blind, one has common SUSY breaking parameters $m_{1/2}$, $m_0$ and $A_0$, representing gaugino and scalar masses and the triliniar coupling. Together with tanβ and the sign of the higgsino mass parameter μ one has four and half parameters at the GUT scale. The magnitude of μ is fixed by the radiative electroweak symmetry breaking condition, discussed below. All the SUSY masses at the weak scale are given in terms of these parameters by the renormalization group evolution (RGE) equations. In particular the gaugino masses evolve like the corresponding gauge couplings at one-loop RGE, so that the bino and wino masses at the weak scale are given by

$$\tilde{B} : M_1 = (\alpha_1/\alpha_G)m_{1/2} \approx 0.4 m_{1/2}; \tilde{W} : M_2 = (\alpha_2/\alpha_G)m_{1/2} \approx 0.8 m_{1/2}. \tag{4}$$

Thus the bino mass is about half the wino mass in this model, so that the latter cannot be the LSP. We can have bino or higgsino LSP dark matter. To see the latter we consider the Higgs scalar masses at the weak scale. These are related to the higgsino mass parameter µ by the radiative symmetry breaking condition

$$\mu^2 + M_Z^2/2 = \frac{M_{Hd}^2 - M_{Hu}^2 \tan^2\beta}{\tan^2\beta - 1} \approx -M_{Hu}^2, \tag{5}$$

where the last equality follows at tanβ > 5, favoured by the large electron-positron (LEP) collider data [6]. Thus the radiative electroweak symmetry breaking is triggered by the $M_{Hu}^2$ changing sign via RGE. For simplicity we shall assume $A_0 = 0$ and positive sign of µ for the time being, so that we have only three input parameters $m_{1/2}$, $m_0$ and tanβ for the RGE. It gives

$$-M_{Hu}^2 = C_1(\alpha_i, h_t, \tan\beta)m_0^2 + C_2(\alpha_i, h_t, \tan\beta)m_{1/2}^2, \tag{6}$$

where $h_t$ denotes the top Yukawa coupling. The coefficient $C_2 \approx 2$, while $C_1 = \pm\varepsilon$ due to an approximate cancellation of the starting value with the top Yukawa contribution to the RGE. For the above mentioned region of tanβ > 5, the sign of this small coefficient is negative ($C_1 = -\varepsilon$). This is called the hyperbolic branch of µ because it is related to $m_0$ and $m_{1/2}$ via the hyperbolic eqs.(5,6) [7]. It is evident from eqs. (4-6) that over most of the parameter space $|\mu|$ is > $M_1$, implying a bino LSP dark matter. However, there is a narrow range of parameters,

$$m_0 \gg m_{1/2} \Rightarrow |\mu| \leq M_1, \tag{7}$$

corresponding to a mixed bino-higgsino or even a dominantly higgsino LSP. The former is referred to as the focus point region [5], while the hyperbolic branch refers to both of them [7]. Note that eq. (7) requires a near cancellation between the two terms of eq. (6), which implies some fine-tuning of these mass parameters for obtaining a mixed bino-higgsino or a dominantly higgsino LSP dark matter.

Fig 1 shows a decade old plot of the $m_0$–$m_{1/2}$ parameter space of the mSUGRA model, going up to the TeV range, for a representative value of tanβ = 10 [8]. The region I is disallowed because of no electroweak symmetry breaking ($\mu^2 < 0$), while the region II is disallowed because of stau $\tilde{\tau}$ LSP. The allowed region is mainly the bino LSP region. Since the bino does not carry any gauge charge, it cannot annihilate via gauge boson exchange, which leads to a generic overabundance of dark matter, as indicated by the relic density contours in the figure. Only the red band near the boundary gives cosmologically compatible dark matter relic density. However, most of this region is disallowed by the Higgs mass bound of $m_h > 115$ GeV from LEP [6], as indicated by the dashed line. In particular it disallows the (fat) bulk annihilation region seen near the origin, where the bino pair annihilate via sfermion exchange,

$$\tilde{B}\tilde{B} \xrightarrow{\tilde{f}} \bar{f}f . \tag{8}$$

This is because the Higgs mass bound from LEP implies $m_{1/2} > 400$ GeV ($M_1 > 2M_Z$), which in turn implies large sfermion masses via RGE, making the annihilation process (8) inefficient. The large sfermion masses are required to suppress the negative loop contribution to $m_h$ from the sfermion partner of top (stop), so that the positive contribution from the top quark loop can raise it from the tree-level limit of $m_h < M_Z \cos 2\beta$ to 115 GeV.

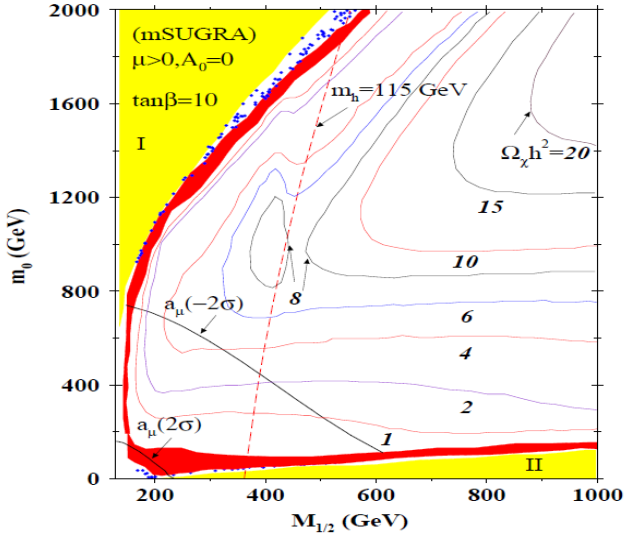

Fig 1. The $m_0$ - $m_{1/2}$ parameter space of mSUGRA model [8]. The cosmologically compatible dark matter relic density region is indicated by the red band, while the blue dots indicate underabundance. The Higgs mass bound from LEP is indicated along with the anomalous muon magnetic moment constraint.

The only cosmologically compatible dark matter relic density region, allowed by the LEP Higgs mass bound, is the stau coannihilation region near the lower boundary, where the bino dark matter coannihilates with a nearly degenerate stau,

$$\tilde{\tau}\tilde{B} \xrightarrow{\tau} \tau\gamma. \tag{9}$$

It requires degeneracy between the stau and bino masses within 10-15 %, which implies some fine-tuning.

The discovery of Higgs boson at $m_h \approx 125$ GeV [9] has pushed up the lower bounds on $m_0$ and $m_{1/2}$ to the TeV region. Therefore the $m_0$–$m_{1/2}$ parameter space of the mSUGRA model must be extended to the multi-TeV range. Fig 2 shows them for two representative values of tanβ = 10 and 50 [10]. These plots use the two-loop RGE code SuSpect [11], which is based the MS-bar renormalization scheme. This scheme is known to give a lower Higgs mass compared to the on-shell renormalization scheme [12] by 2-3 GeV. Therefore we show the predicted Higgs mass band of 122-125 GeV in this figure.

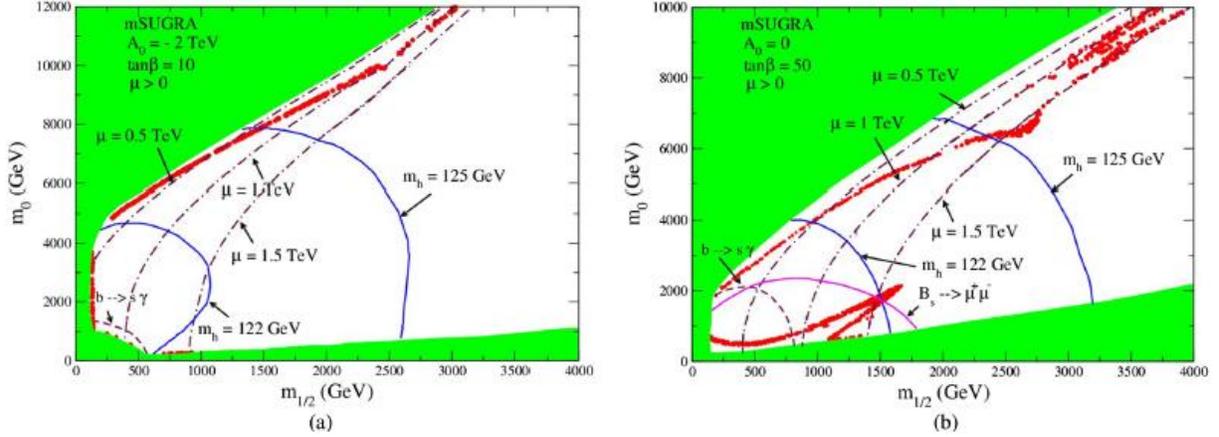

Fig 2. The $m_0 - m_{1/2}$ parameter space of mSUGRA model, extended into the multi-TeV range, at tenβ = 10 (a) and 50 (b) [10]. The red dots indicate the regions compatible with the cosmic dark matter relic density measurement.

The red dots indicate the regions compatible with the cosmic dark matter relic density measurement,

$$0.09 < \Omega_\chi h^2 < 0.14, \tag{10}$$

corresponding to 5σ range of the Plank data [13], which also accommodates the WMAP data [14].

One can see the following four dark matter relic density compatible regions in this figure.

1. The stau coannihilation region is indicated by a few red dots near the lower boundary of the left figure. It involves some fine-tuning between the co-annihilating stau and bino masses, as mentioned above. It is also seen to be seriously challenged by the Higgs mass constraint from LHC. This constraint is in fact stronger than that coming from SUSY signal search at LHC.
2. The so called funnel region, occurring near the bottom of the right figure, corresponds to resonant annihilation of the LSP pair into a pair of fermions via the pseudo scalar Higgs boson,

$$\chi\chi \xrightarrow{A} \bar{f}f \,. \tag{11}$$

   Since the Higgs boson couplings to the $\chi$ are proportional to the product of its gaugino and higgsino components in (1),

$$g_{A\chi\chi}, g_{H\chi\chi} \propto c_{1,2} c_{3,4} \,, \tag{12}$$

   they are suppressed for a bino dominated $\chi$. To compensate for this one needs a Breit-Wigner enhancement from the resonant annihilation condition,

$$M_A \approx 2M_1 \,, \tag{13}$$

   which implies some fine-tuning. One also needs to go to large tanβ to enhance the A coupling to the fermions. One clearly sees from the right figure that this region is seriously challenged by the Higgs mass constraint [9] as well as the $B_s \to \mu\mu$ decay constraint [15, 16] from the LHC.
3. The focus point region corresponds to the left half of the red strip near the upper boundary of both the figures, where

   $\mu \sim M_1 \sim 0.5$ TeV. (14)

   It implies a mixed bino – higgsino LSP, i.e. significant higgsino components $c_{3,4}$ of $\chi$ in (1). Since the Z boson coupling to $\chi$ is proportional to the square of its higgsino components,

$$g_{Z\chi\chi} \propto c_3^2 - c_4^2 \,, \tag{15}$$

   the $\chi$ pair can efficiently annihilate via Z boson, i.e.

$$\chi\chi \xrightarrow{Z} \bar{f}f \,. \tag{16}$$

   As mentioned earlier, the condition (14) requires an approximate cancellation between the two large terms of eq. (6), implying some fine-tuning. The mixed bino-higgsino LSP also has large coupling to the Higgs bosons (12), which implies a large direct detection cross-section of $\chi$ via Higgs exchange. In fact this region is seriously challenged by the negative results from the direct dark matter detection experiments, as shown in ref [17,18].

4. Finally, the right end of the upper red band in both the figures corresponds to the higgsino LSP region,

$$M_{\tilde{H}^{\pm,0}} = \mu \approx 1 TeV. \quad (17)$$

Note that the higgsino LSP condition, $\mu < M_1$, implies even a closer cancellation between the two terms of eq. (6), implying even a larger fine-tuning than the focus point region. Since higgsinos carry isospin, $I = ½$, they can annihilate via their isospin gauge coupling to the W boson, i.e.

$$\tilde{H}^{\pm}\tilde{H}^0 \xrightarrow{W} \bar{f}f. \quad (18)$$

Since the annihilation occurs via gauge coupling, its rate does not depend on any SUSY parameter other than the higgsino mass; and the cosmic relic density of eq. (10) corresponds to higgsino dark matter mass of about 1 TeV. This is a robust result that holds in any higgsino LSP model. In the universal model, the higgsino LSP condition, $M_1 > 1$ TeV, implies

$$M_3 = (\alpha_3/\alpha_1)M_1 > 7 TeV, \quad (19)$$

i.e. squarks and gluinos heavier than 7 TeV, which is beyond the reach of LHC. Besides, a 1 TeV higgsino LSP offers no viable signal for direct dark matter detection. It was shown in [19], that a 1 TeV higgsino LSP signal can be seen at a 3 TeV electron-positron collider (CLIC) via anomalous single photon events. It is widely believed, however, that there will be no CLIC if there is no SUSY signal at LHC. So the higgsino LSP model may have little relevance to collider phenomenology, at least in the CMSSM. We shall see later that one expects a more favourable collider signal of higgsino LSP in some nonuniversal gaugino mass models.

In summary, all the four dark matter relic density compatible regions of CMSSM suffer from some amount of fine-tuning. Besides three of them are seriously challenged by the constraints from LHC and direct dark matter detection experiments, while the fourth one is out of reach for both of them. We shall see in the following sections that some of the nonuniversal gaugino mass models offer more promising dark matter candidates for collider and dark matter detection experiments.

## 3. Nonuniversality of Gaugino Masses in SU(5) GUT:

The gauge kinetic function responsible for the gaugino masses in the GUT scale Lagrangian originates from the vacuum expectation value of the F term of a chiral superfield $\Phi$, which is responsible for SUSY breaking, i.e.

$$L = \frac{<F_\Phi>_{ij}}{M_{Planck}} \lambda_i \lambda_j \quad . \tag{20}$$

Here $\lambda_{1,2,3}$ are the U(1), SU(2), SU(3) gaugino fields – bino, wino and gluino. Since gauginos belong to the adjoint representation of the GUT group, $\Phi$ and $F_\Phi$ can belong to any of the irreducible representations occurring in their symmetric product, i.e.

$$24 \times 24 = 1 + 24 + 75 + 200. \tag{21}$$

Thus the GUT scale gaugino masses for a given representation of the SUSY breaking superfield are determined in terms of one mass parameter as

$$M^G_{1,2,3} = C^n_{1,2,3} m_{1/2}, \tag{22}$$

where the values of the coefficients $C^n_{1,2,3}$ are listed in Table 1 [20]. The coefficients $C^n_3$ are conventionally normalized to 1.

| $n$ | $M^G_3$ | $M^G_2$ | $M^G_1$ |
|---|---|---|---|
| 1 | 1 | 1 | 1 |
| 24 | 1 | $-3/2$ | $-1/2$ |
| 75 | 1 | 3 | $-5$ |
| 200 | 1 | 2 | 10 |

Table 1: Relative values of the SU(3), SU(2) and U(1) gaugino masses at GUT scale for different representations $n$ of the chiral superfield $\Phi$.

The CMSSM assumes the SUSY breaking superfield $\Phi$ to be a singlet, implying universal gaugino masses at the GUT scale. On the other hand, any of the three nonsinglet representations of $\Phi$ would imply nonuniversal gaugino masses as per Table 1 [20]. These nonuniversal gaugino mass models are known to be consistent with the universality of gauge couplings, $\alpha_G \approx 1/25$, and their phenomenology have been widely studied [21]. The superparticle masses at the weak scale are related to these GUT scale gaugino masses along with the universal scalar mass and trilinear coupling parameter, m₀ and A₀, via the RGE. In particular the gaugino masses evolve like the corresponding gauge couplings at the one-loop level of the RGE (4), implying

$$
\begin{aligned}
M_1 &= (\alpha_1/\alpha_G)M_1^G \approx (25/60)C_1^n m_{1/2}, \\
M_2 &= (\alpha_2/\alpha_G)M_2^G \approx (25/30)C_2^n m_{1/2}, \\
M_3 &= (\alpha_3/\alpha_G)M_3^G \approx (25/9)m_{1/2}.
\end{aligned}
\qquad (23)
$$

The corresponding higgsino mass μ is obtained from the electroweak symmetry breaking condition along with the RGE for the Higgs scalar masses. If one neglects the A₀ contributions, then one has a relatively simple expression for the higgsino mass at the one-loop level of the RGE [22], i.e.

$$
\begin{aligned}
\mu^2 + \frac{1}{2}M_Z^2 \simeq\ & -0.1 m_0^2 + 2.1 M_3^{G2} - 0.22 M_2^{G2} - 0.006 M_1^{G2} + 0.006 M_1^G M_2^G + \\
& 0.19 M_2^G M_3^G + 0.03 M_1^G M_3^G,
\end{aligned}
\qquad (24)
$$

where the numerical coefficients on the right hand side correspond to a representative value of tan β = 10, but have only modest variations over the moderate tan β region.

Our results are based on exact numerical solutions to the two-loop RGE including the A₀ contributions using the SuSpect code [11]. Nonetheless, the approximate formulae (23) and (24) are very useful in understanding the composition of the LSP dark matter in these models. The dominant contribution to the higgsino mass formula (24) comes from the $M_3^G$ term, implying μ ~ √2 $m_{1/2}$ from Table 1 for all the four models. On the other hand the bino mass formula (23) shows that $M_1$ ~ 0.4$m_{1/2}$ in the CMSSM, implying a bino dominated LSP dark matter in this model as seen in the last section. One sees from Table 1 that $M_1$ is further suppressed by a factor of half in the 24 model, implying an even more strongly bino dominated LSP dark matter. Thus one predicts a generic overabundance of dark matter relic density in the CMSSM as well as the 24 model. For the 75 and the 200 models, however, one sees from table 1 that the bino mass $M_1$ is enhanced by factors 5 and 10 respectively relative to the CMSSM, implying a higgsino dominated LSP dark matter in these nonuniversal gaugino mass models. Since higgsino has an efficient annihilation mechanism via its isospin gauge coupling to W boson, one gets the cosmologically compatible dark matter relic density in these models for a higgsino mass μ ~ 1 TeV, as we had seen at the end of last section. We shall discuss the phenomenology of higgsino dark matter in the 75 and 200 models in section 6. Before that we shall consider the phenomenology of nonuiversal gaugino mass models (NUGM), based on SUSY breaking via a combination of a singlet and a nonsinglet superfield. In section 4 we shall consider a combination of a dominant singlet with a subdominant nonsinglet superfield, so that the LSP dark matter remains dominantly bino. But in these models one can get cosmologically compatible dark matter relic density naturally via bulk annihilation, while satisfying the Higgs mass and other collider constraints. In section 5 we shall consider SUSY breaking models with comparable singlet and nonsinglet components, corresponding to a mixed bino-higgsino LSP dark matter.

## 4. Bulk Annihilation Region of Bino DM in the 1+75 and 1+200 Models:

Nonuniversal gaugino mass models for bino dark matter were considered in [23], assuming SUSY breaking via the combination of a singlet and a nonsinglet superfield, i.e.

$$\langle F \rangle = (1-\alpha)\langle F_1 \rangle + \alpha \langle F_{24,75,200} \rangle, \qquad (25)$$

where $\langle F \rangle$ represents the vev of the F term of the SUSY breaking superfield. In these models one could reconcile the bulk annihilation region of cosmic dark matter relic density with the Higgs and other mass limits from LEP. Fig 3 shows the parameter space of nonuniversal GUT scale gugino masses $M_{1,2}^G$ for a fixed value of $M_3^G = 600$ GeV at $m_0$ = 70 GeV, tanβ = 10 and $A_0$ = 0. The areas excluded by the wrong LSP, no radiative electroweak symmetry breaking, and the LEP mass limits are indicated by the various shaded areas. The only allowed regions are the elliptic regions, of which the elliptic bands are compatible with the cosmic dark matter relic density. The fine-tuning parameter for this band is colour coded, the most natural region being the yellow region corresponding to bulk annihilation.

The three straight lines show the parameter space spanned by the 1+24, 1+75 and 1+200 models. Their meeting point represents the universal model (CMSSM), α = 0, where the three GUT scale gaugino masses have a common value of 600 GeV.

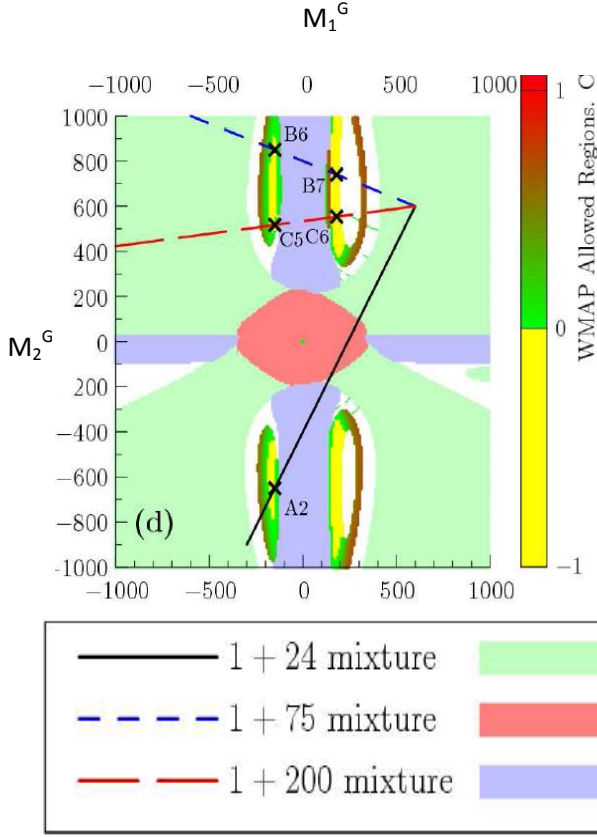

Fig 3. The GUT scale gaugino mass $M_1^G - M_2^G$ plot of NUGM models at fixed $M_3^G = 600$ GeV, $m_0 = 70$ GeV and $\tan\beta = 10$. The cosmic DM relic density compatible regions are the elliptic bands with fine-tuning colour coding. The yellow band (no fine-tuning) corresponds to the bulk annihilation region [23].

We see from Fig 3 that the bulk annihilation region of cosmic dark matter relic density can be reached in the 1+75 and 1+200 models with dominant singlet contributions (small α) and in the 1+24 model with a dominant nonsinglet contribution (α ≈ 1). The relatively large value of $M_3^G = 600$ GeV implies heavy gluino and squarks in the TeV range, raising Higgs mass above the LEP limit of 115 GeV. On the other hand, the small values of $m_0$ and $M_1^G$ ensure relatively light bino LSP and right sleptons (SUSY partners of right-handed leptons) in the mass range of ~ 100 GeV, which ensure efficient bulk annihilation of bino pair (8) via right slepton exchange.

The 1+75 and 1+200 models were revisited in [24], after the discovery of the Higgs boson. Assuming a TeV scale trilinear coupling parameter $A_0$, one gets an additional radiative contribution to $m_h$ from left-right stop mixing, which pushes it up to the desired range of 122 – 125 GeV. Thus one can reconcile the Higgs mass constraint from LHC with the bulk annihilation region of cosmic dark matter relic density in these NUGM models. Table 2 shows the weak scale superparticle masses for two representative points of these two models at $\tan\beta = 10$ and a large $A_0 = -1300$ GeV.

Table2. The superparticle masses for two representative points of the 1+75 and 1+200 models at tanβ = 10 and $A_0$ = -1300 GeV [24].

$M_1^G = 250$ GeV, $M_3^G = 800$ GeV, $m_0 = 80$ GeV

| Particle | Mass (GeV) | |
|---|---|---|
| | (1+75) model | (1+200) model |
| $\tilde{\chi}_1^0$ (bino) | 101 | 101 |
| $\tilde{\chi}_2^0$ (wino) | 789 | 593 |
| $\tilde{\chi}_3^0$ (higgsino) | 1197 | 1218 |
| $\tilde{\chi}_4^0$ (higgsino) | 1206 | 1223 |
| $\tilde{\chi}_1^+$ (wino) | 789 | 592 |
| $\tilde{\chi}_2^+$ (higgsino) | 1206 | 1223 |
| $M_1$ | 103 | 103 |
| $M_2$ | 780 | 581 |
| $M_3$ | 1728 | 1732 |
| $\mu$ | 1197 | 1217 |
| $\tilde{g}$ | 1766 | 1767 |
| $\tilde{\tau}_1$ | 109 | 111 |
| $\tilde{\tau}_2$ | 649 | 478 |
| $\tilde{e}_R, \tilde{\mu}_R$ | 128 | 128 |
| $\tilde{e}_L, \tilde{\mu}_L$ | 631 | 477 |
| $\tilde{t}_1$ | 1056 | 1096 |
| $\tilde{t}_2$ | 1488 | 1455 |
| $\tilde{b}_1$ | 1459 | 1421 |
| $\tilde{b}_2$ | 1519 | 1524 |
| $\tilde{q}_{1,2,R}$ | ~1531 | ~1536 |
| $\tilde{q}_{1,2,L}$ | ~1643 | ~1597 |

One can clearly see the coexistence of TeV scale squark and gluino masses along with relatively light bino LSP and right sleptons in the mass range of ~ 100 GeV in these models. The former ensures compatibility with the Higgs mass and direct SUSY search constraints from LHC, while the latter ensures cosmologically compatible dark matter relic density via bulk annihilation.

With a bino dominated LSP dark matter these models predict rather small direct detection cross-sections as expected from (12). We see from Fig 4 [24] that the predicted cross-sections are compatible with the limits from the present direct detection experiments [25,26], but within the reach of the 1 Ton XENON proposal. These cross-sections were computed using the DarkSUSY code [27].

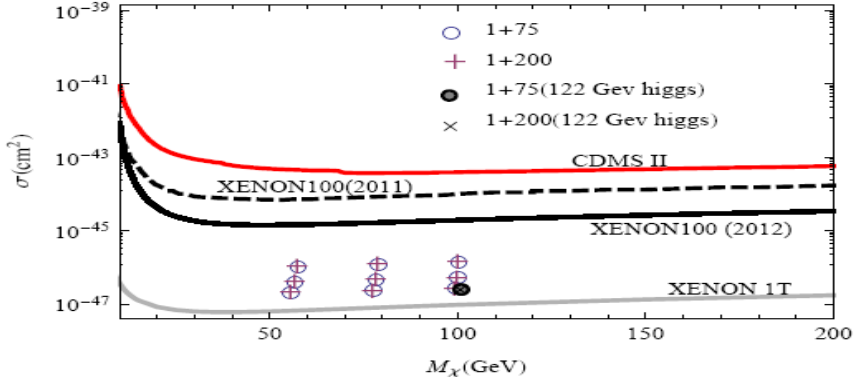

Fig 4. Direct dark matter detection cross-sections of the 1+75 and 1+200 models compared with the limits of the present experiments and the reach of the 1 Ton XENON proposal [24].

Interestingly, the radiative pair annihilation process,

$$\tilde{\chi}_1^0 + \tilde{\chi}_1^0 \xrightarrow{\tilde{e}} e^+ + e^- + \gamma, \tag{26}$$

can give a hard positron signal of the type observed by the PAMELA experiment [28]. Unfortunately the size of the predicted signal is smaller by a factor of ten thousand compared to the data.

Going back to the Table 2, we see that the 1+200 model has relatively low wino and left slepton masses compared to the 1+75 model. Therefore it offers the interesting possibility of explaining the observed anomalous magnetic moment of muon [29],

$$\Delta a_\mu = (2.87 \pm 0.80) \times 10^{-9}, \tag{27}$$

via the SUSY loop contribution of Fig 5 [30].

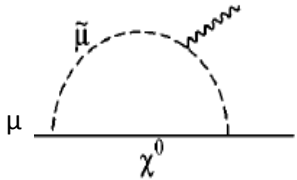

Fig 5. SUSY loop contribution to the anomalous magnetic moment of muon.

Table 3. The superparticle and Higgs masses for some representative points of the 1+200 model shown along with the contribution to $a_\mu$ at $\tan\beta = 15$, $m_0=103$ GeV and $A_0= -1400$ GeV [31].

| Particle | All masses in GeV | | | | |
|---|---|---|---|---|---|
| | $M_1^G = 220$ | $M_1^G = 200$ | | | |
| | $M_3^G = 600$ | $M_3^G = 600$ | $M_3^G = 700$ | $M_3^G = 800$ | $M_3^G = 900$ |
| $\tilde{\chi}_1^0$ (bino) | 88.8 | 80.2 | 79.9 | 79.6 | 79.1 |
| $\tilde{\chi}_2^0$ (wino) | 445 | 443 | 516 | 589 | 662 |
| $\tilde{\chi}_3^0$ (higgsino) | 1032 | 1031 | 1132 | 1233 | 1333 |
| $\tilde{\chi}_4^0$ (higgsino) | 1036 | 1036 | 1137 | 1237 | 1337 |
| $\tilde{\chi}_1^+$ (wino) | 445 | 443 | 516 | 589 | 662 |
| $\tilde{\chi}_2^+$ (higgsino) | 1036 | 1037 | 1137 | 1237 | 1337 |
| $M_1$ | 90.1 | 81.4 | 81.1 | 80.8 | 80.4 |
| $M_2$ | 437 | 435 | 506 | 577 | 648 |
| $M_3$ | 1331 | 1331 | 1533 | 1734 | 1933 |
| $\mu$ | 1034 | 1033 | 1133 | 1233 | 1332 |
| $\tilde{g}$ | 1354 | 1354 | 1561 | 1766 | 1969 |
| $\tilde{\tau}_1$ | 103 | 97.4 | 96.2 | 93.6 | 89.8 |
| $\tilde{\tau}_2$ | 381 | 379 | 429 | 480 | 532 |
| $\tilde{e}_R, \tilde{\mu}_R$ | 138 | 134 | 133 | 133 | 132 |
| $\tilde{e}_L, \tilde{\mu}_L$ | 374 | 373 | 425 | 477 | 530 |
| $\tilde{t}_1$ | 748 | 749 | 920 | 1082 | 1240 |
| $\tilde{t}_2$ | 1107 | 1107 | 1275 | 1441 | 1607 |
| $\tilde{b}_1$ | 1056 | 1055 | 1232 | 1405 | 1576 |
| $\tilde{b}_2$ | 1161 | 1161 | 1336 | 1509 | 1681 |
| $\tilde{q}_{1,2,R}$ | $\sim 1188$ | $\sim 1188$ | $\sim 1364$ | $\sim 1538$ | $\sim 1710$ |
| $\tilde{q}_{1,2,L}$ | $\sim 1233$ | $\sim 1233$ | $\sim 1417$ | $\sim 1598$ | $\sim 1778$ |
| $h$ | 122 | 122 | 122 | 122 | 123 |
| Muon $g-2$ | | | | | |
| $a_\mu$ | $2.22 \times 10^{-9}$ | $2.28 \times 10^{-9}$ | $1.89 \times 10^{-9}$ | $1.59 \times 10^{-9}$ | $1.37 \times 10^{-9}$ |
| $(\delta a_\mu)$ | $(0.83\sigma)$ | $(0.75\sigma)$ | $(1.24\sigma)$ | $(1.61\sigma)$ | $(1.89\sigma)$ |

Table 3 shows the superparticle and Higgs boson masses for some representative points of the 1+200 model along with the SUSY contribution to muon anomalous magnetic moment [31], computed using the micrOMEGA code [32]. We see that one can explain the anomalous $a_\mu$ to within $2\sigma$ with a modest wino mass in the range of 400-700 GeV. Note that the left sleptons are always lighter than the wino. Therefore we can probe this SUSY model via electroweak production of wino pairs at LHC, followed by their leptonic decays via the left sleptons, i.e.

$$\bar{q}q \to \tilde{\chi}_1^+ \tilde{\chi}_2^0, \tilde{\chi}_1^+ \xrightarrow{\tilde{l}_L} l^+ \nu \tilde{\chi}_1^0, \tilde{\chi}_2^0 \xrightarrow{\tilde{l}_L} l^+ l^- \tilde{\chi}_1^0. \qquad (28)$$

A simulation study of this process shows promising same sign dilepton and trilepton + large missing-$E_T$ signals without any accompanying hard jets [33]. It should be possible to probe the entire wino mass range up to 700 GeV, which can account for the anomalous $a_\mu$ within 2σ, via these signals with the current LHC run.

It should be noted, however, that the anomalous magnetic moment of muon is not a compelling constraint on SUSY models, because it is only a 3σ anomaly; and besides there are viable alternative models for this anomaly based e.g. on extra gauge bosons associated with the $L_\mu$-$L_\tau$ symmetry [34]. Therefore one needs to probe also the NUGM models like the 1+75, which are compatible with cosmic dark matter relic density and the collider results, but predict relatively heavy wino that cannot be probed via the elctroweak production process (28). They can be probed via gluino/squark pair production at LHC, followed by their cascade decays, resulting in same sign dilepton and trilepton + large missing-$E_T$ signals that are accompanied by hard hadronic jets. Signals of this type have been suggested in the past for many versions of MSSM [35]. But the presence of left sleptons, ligter than wino, promises large leptonic signals for these SUSY models.

## 5. Mixed Gaugino-Higgsino DM in the 1+75 and 1+200 Models:

In this section we consider NUGM models based on SUSY breaking by a comparable admixture of singlet and nonsinglet superfields. The resulting LSP dark matter is a mixed bino-higgsino state for the 1+75 model and a triply mixed bino-wino-higgsino state for the 1+200 model [36]. These are examples of the well-tempered neutralino scenario [4], mentioned in the introduction. They are compatible with cosmic dark matter relic density over wide regions of the parameter space. Being mixed gaugino-higgsino states, the dark matter pair can efficiently annihilate via Z boson (16) and also via the pseudo scalar Higgs boson (11). The former is analogous to the focus point region and the latter to the funnel region of the CMSSM. However, cosmologically compatible dark matter relic density is achieved over much broader bands of parameter space compared to the CMSSM. Moreover the funnel region is achieved also at low tanβ , as we see below.

Fig 6 shows the $m_0$–$m_{1/2}$ parameter space of the 1+75 model at $\tan\beta = 10$ [36]. The WMAP compatible dark matter relic density regions are shown by the red bands, computed using the micrOMEGA code [32].

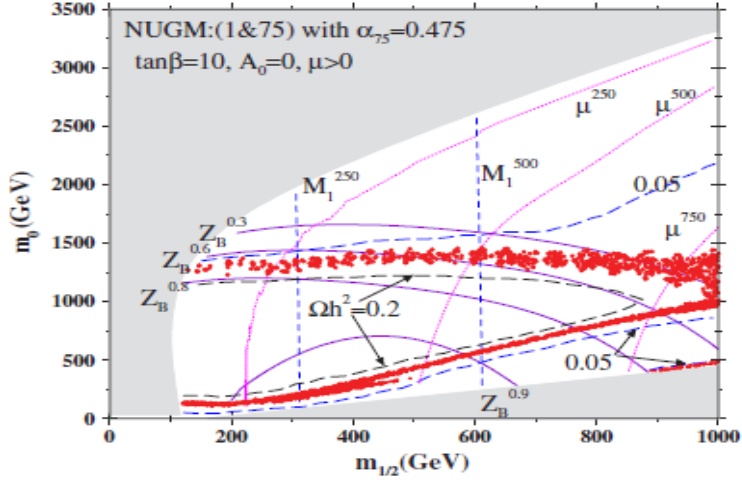

Fig 6. The $m_0$-$m_{1/2}$ parameter space of the 1+75 model, with the WMAP dark matter relic density compatible region shown as the red bands. The bino content of dark matter is indicated by the $Z_B$ = 0.3, 0.6, 0.8 and 0.9 contours [36].

The upper band corresponds to mixed bino-higgsino dark matter, pair annihilating via Z boson (16). The lower band corresponds to a relatively dominant bino dark matter, pair annihilating via pseudoscalar Higgs boson (11), analogous to the funnel region. Note the underabundance of dark matter relic density between the two arms of the funnel region and above the upper band, while there is an overabundance (>0.2) between the two bands.

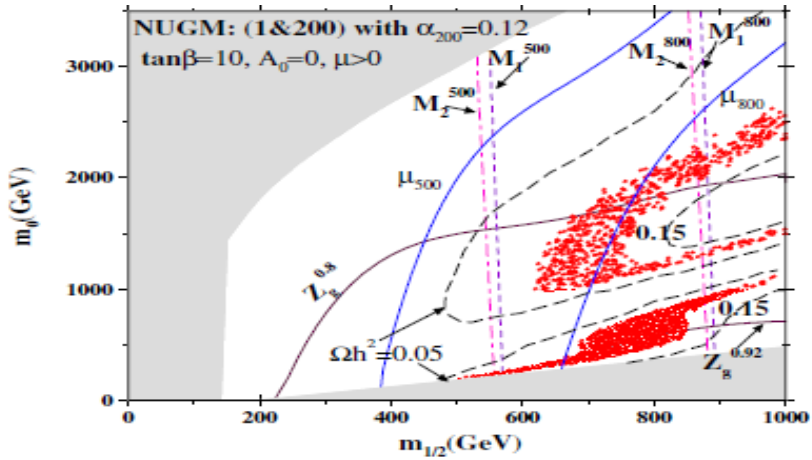

Fig 7. The $m_0$-$m_{1/2}$ parameter space for the 1+200 model [36].

Fig 7 shows the analogous plot for the 1+200 model, with a mixed bino-wino-higgsino dark matter [36]. The $Z_g$ contours show the gaugino content of dark matter. Again the upper band corresponds to a mixed gaugino-higgsino dark matter, pair annihilating via Z boson (16). The two lower bands represent the two arms of the funnel region, where the dark matter pair annihilate via pseudoscalr Higgs boson (11). A simulation study shows promising signatures of these models at LHC [37]. However, one expects even a more promising signature of these models in direct dark matter detection experiments, since the signal cross-section via Higgs boson exchange (12) is expected to be large for a mixed gaugino-higgsino state. Indeed the two candidate events reported by the CDMS experiment [25] were shown to be in agreement with the prediction of the mixed gaugino-higgsino regions (upper bands) of these models [38].

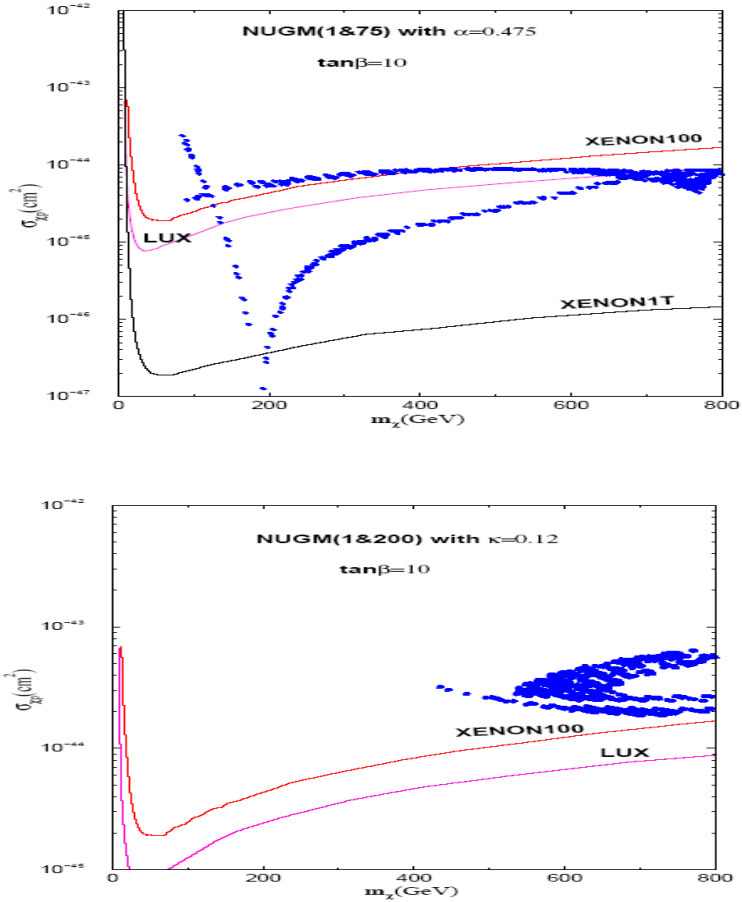

Fig 8. The 1+75 and 1+200 model predictions for direct dark matter detection cross-section compared with the limits from XENON and LUX experiments [26, 39, 40].

Unfortunately, the CDMS result has been superseded by the negative results from two more sensitive experiments, XENON 100 [26] and LUX [39]. Fig 8 compares the predictions of the WMAP compatible dark matter relic density regions (blue dots) of the 1+75 and 1+200 models with these experimental limits. They seem to rule out the mixed gaugino-higgsino dark matter region of the 1+75 model, represented by the upper band, and the entire WMAP compatible region of the 1+200 model. Thus like the focus point region of the CMSSM, these mixed dark matter models are also strongly disfavoured by the direct dark matter detection data.

## 6. Higgsino DM in the 75 and 200 Models:

While discussing the higgsino dark matter of the CMSSM, we had mentioned that the cosmologically compatible dark matter relic density is achieved for a higgsino mass of ~ 1 TeV irrespective of the SUSY model. However the higgsino dark matter of the 75 and 200 models have two major advantages over that of the CMSSM. In these models the higgsino LSP dark matter occurs naturally without any fine-tuning of the $m_0$ and $m_{1/2}$ parameters. Moreover, the cosmologically compatible dark matter relic density in these models is achieved for relatively lower gluino and squark mass region, at least a part of which can be probed at the LHC.

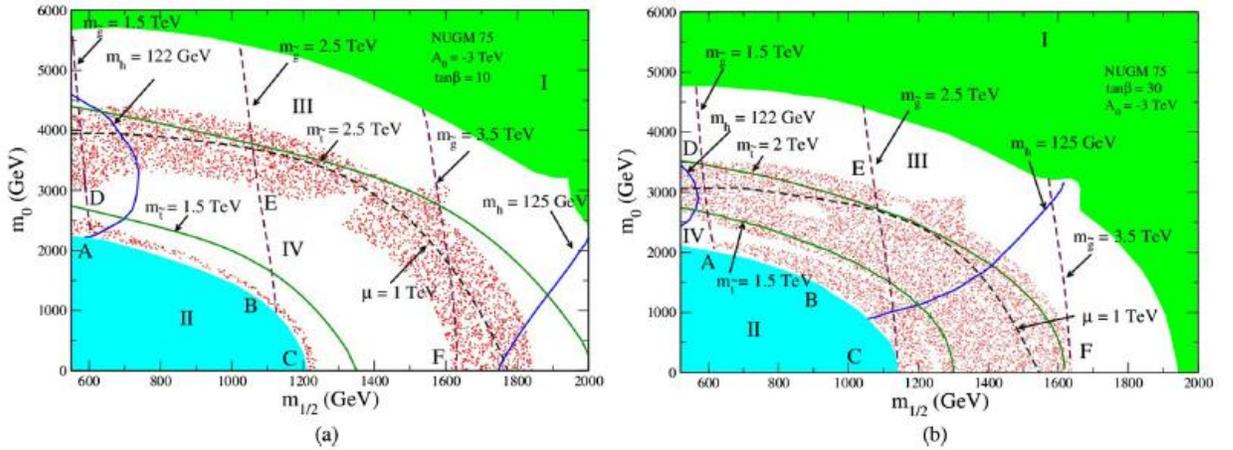

Fig 9. The $m_0$-$m_{1/2}$ plot of the 75 model for two representative values of tanβ = 10 (a) and 30 (b) [10]. The red dotted bands satisfy the cosmologically compatible dark matter relic density (10), while regions III and IV represent under abundance and overabundance respectively. They are computed using the micrOMEGA code [32].

Fig 9 shows the $m_0$-$m_{1/2}$ plot of the 75 model for two representative values of tanβ = 10 and 30 [10]. The regions marked I and II are disallowed because they correspond to no electroweak symmetry breaking and to stop LSP respectively. The red dotted bands around the µ = 1 TeV contour, marked DEF, satisfy cosmologically compatible dark matter relic density (10) due to coannihilation of changed and neutral higgsinos (18). The bands adjacent to the lower boundary, marked ABC, also get large contribution from coannihilation of higgsino with stop. The dashed vertical lines represent gluino mass of 1.5, 2.5 and 3.5 TeV, while the blue solid lines correspond to Higgs mass of 122 and 125 GeV. The green contours represent stop mass of 1.5 and 2.5 (2.0) TeV for the left (right) figure. We see that at least a part of the cosmologically compatible higgsino dark matter region corresponds to gluino and stop masses ≤ 2 TeV, which can be probed at the LHC. Table 4 lists the SUSY masses for some benchmark points of this region.

Table 4. The SUSY masses for some benchmark points of Fig 9 corresponding to gluino and stop masses ≤ 2 TeV [10].

| Parameter | 1 | 2 | 3 | 4 | 5 | 6 |
|---|---|---|---|---|---|---|
| $m_{1/2}$ | 837.11 | 731.80 | 658.32 | 843.10 | 762.16 | 656.10 |
| $m_0$ | 1948.54 | 3689.09 | 2248.34 | 1805.52 | 3110.81 | 2089.04 |
| $\tan\beta$ | 10.00 | 10.00 | 10.00 | 30.00 | 30.00 | 30.00 |
| $A_0$ | −3000.00 | −3000.00 | −3000.00 | −3000.00 | −3000.00 | −3000.00 |
| $(M_1, M_2, M_3)$ | 1884, 2035. 1767 | 1661, 1786, 1517 | 1475, 1599, 1400 | 1895, 2052, 1786 | 1724, 1860, 1593 | 1467, 1595, 1403 |
| $\mu$ | 1268.56 | 1062.37 | 1235.64 | 1176.45 | 978.18 | 1148.12 |
| $m_{\tilde{g}}$ | 1954.41 | 1822.47 | 1600.97 | 1957.89 | 1861.66 | 1587.32 |
| $m_{\tilde{\chi}_1^\pm}, m_{\tilde{\chi}_2^\pm}$ | 1272.51, 2081.25 | 1072.96, 1854.31 | 1234.37, 1650.76 | 1181.44, 2095.35 | 988.12, 1922.20 | 1148.78, 1643.44 |
| $m_{\tilde{\chi}_1^0}, m_{\tilde{\chi}_2^0}$ | 1272.18, 1275.75 | 1072.59, 1076.29 | 1233.99, 1240.02 | 1180.92, 1183.69 | 987.55, 990.32 | 1148.18, 1152.51 |
| $m_{\tilde{t}_1}, m_{\tilde{t}_2}$ | 1298.52, 2421.22 | 2188.85, 3365.72 | 1282.69, 2287.72 | 1270.36, 2235.20 | 1909.76, 2838.59 | 1220.38, 2054.05 |
| $m_{\tilde{b}_1}, m_{\tilde{b}_2}$ | 2412.53, 2486.19 | 3364.03, 3866.90 | 2279.98, 2533.74 | 2098.72, 2229.44 | 2835.95, 3048.80 | 2038.31, 2127.47 |
| $m_{\tilde{u}_1}, m_{\tilde{u}_2}$ | 2922.94, 2671.38 | 4104.89, 3982.31 | 2818.42, 2661.21 | 2848.60, 2584.39 | 3657.64, 3500.96 | 2697.29, 2531.82 |
| $m_{\tilde{e}_L}, m_{\tilde{e}_R}$ | 2620.75, 2480.56 | 3978.62, 3919.23 | 2633.97, 2550.39 | 2527.42, 2377.81 | 3485.48, 3405.96 | 2499.53, 2410.04 |
| $m_{\tilde{\tau}_1}, m_{\tilde{\tau}_2}$ | 2450.19, 2606.89 | 3881.97, 3960.64 | 2519.42, 2619.58 | 2096.37, 2402.76 | 3078.22, 3331.77 | 2122.94, 2368.87 |
| $m_A, m_{H^\pm}$ | 2873.09, 2873.11 | 4068.20, 4068.24 | 2865.53, 2865.33 | 2210.31, 2210.25 | 2913.68, 2913.65 | 2179.37, 2179.33 |
| $m_h$ | 123.13 | 122.04 | 122.20 | 123.95 | 122.87 | 123.03 |
| $\Omega_{\tilde{\chi}_1} h^2$ | 0.11 | 0.13 | 0.12 | 0.12 | 0.11 | 0.10 |
| $BF(b \to s\gamma)$ | $3 \times 10^{-4}$ | $3.04 \times 10^{-4}$ | $3 \times 10^{-4}$ | $2.77 \times 10^{-4}$ | $2.91 \times 10^{-4}$ | $2.75 \times 10^{-4}$ |
| $BF(B_s \to \mu^+\mu^-)$ | $3.53 \times 10^{-9}$ | $3.53 \times 10^{-9}$ | $3.53 \times 10^{-9}$ | $3.76 \times 10^{-9}$ | $3.58 \times 10^{-9}$ | $3.76 \times 10^{-9}$ |
| $\sigma_{p\chi}^{SI}$ in pb | $7 \times 10^{-10}$ | $7.57 \times 10^{-10}$ | $2.97 \times 10^{-9}$ | $4.54 \times 10^{-10}$ | $4.11 \times 10^{-10}$ | $1.74 \times 10^{-9}$ |
| $\sigma_{gg}^{NLO}$ in fb | 1.23 | 2.83 | 8.45 | 1.19 | 2.21 | 8.97 |
| Dominant decay modes of $\tilde{g}$ (in %) (> 10% are shown) | $\tilde{g} \to \tilde{t}_1 \bar{t}$ 50<br>$\to \tilde{t}_1^* t$ 50 | $\tilde{g} \to \tilde{\chi}_1^0 t\bar{t}$ 23<br>$\to \tilde{\chi}_2^0 t\bar{t}$ 18<br>$\to \tilde{\chi}_1^- t\bar{b}$ 27<br>$\to \tilde{\chi}_1^+ b\bar{t}$ 27 | $\tilde{g} \to \tilde{t}_1 \bar{t}$ 50<br>$\to \tilde{t}_1^* t$ 50 | $\tilde{g} \to \tilde{t}_1 \bar{t}$ 50<br>$\to \tilde{t}_1^* t$ 50 | $\tilde{g} \to \tilde{\chi}_1^0 t\bar{t}$ 23<br>$\to \tilde{\chi}_2^0 t\bar{t}$ 20<br>$\to \tilde{\chi}_1^- t\bar{b}$ 27<br>$\to \tilde{\chi}_1^+ b\bar{t}$ 27 | $\tilde{g} \to \tilde{t}_1 \bar{t}$ 50<br>$\to \tilde{t}_1^* t$ 50 |
| Dominant decay modes of $\tilde{t}_1$ (in %) (> 10% are shown) | $\tilde{t}_1 \to b\tilde{\chi}_1^+$ 100 | ... | $\tilde{t}_1 \to b\tilde{\chi}_1^+$ 100 | $\tilde{t}_1 \to b\tilde{\chi}_1^+$ 100 | ... | $\tilde{t}_1 \to b\tilde{\chi}_1^+$ 100 |

One sees from table 4 that the lighter stop $\tilde{t}_1$ is significantly lighter than the other squarks, i.e. an inverted mass hierarchy of squarks compared to the quarks. This along with the higgsino dominance of the degenerate lighter chargino and neutralino states ($\tilde{\chi}_1^+, \tilde{\chi}_{1,2}^0$) leads to the dominance of the gluino decay modes,

$$\tilde{g} \to t\bar{t}\tilde{\chi}_{1,2}^0; \tilde{g} \to t\bar{b}\chi_1^-. \tag{29}$$

Thus one expects the decay of the produced gluino pair at LHC to give a distinctive SUSY signal with two same sign tops or 3-4 tops along with a large missing-$E_T$. From the listed glino pair production cross-sections in fb units, one expects ~ 100 signal events for these benchmark points at the high luminosity LHC run of $L \geq 100$ fb$^{-1}$ [10].

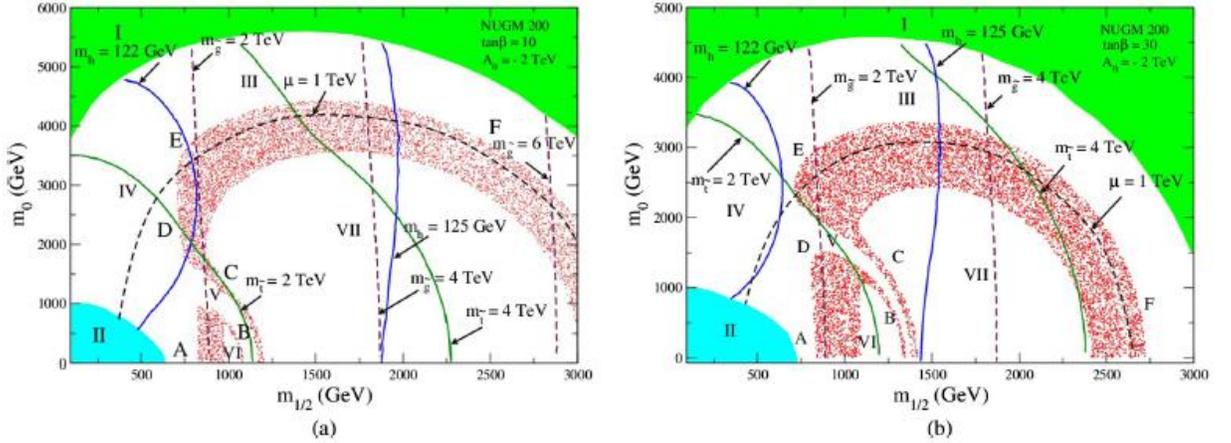

Fig 10. Same as Fig 9 for the 200 model.

Fig 10 shows the analogous $m_0$-$m_{1/2}$ plot for the 200 model, again with the cosmologically compatible dark matter relic density indicated by the band of red dots. Over most of this band, overlapping with the μ = 1 TeV contour, this relic density is achieved by the charged and neutral higgsino coannihilation process (18) via s-channel W boson. In the region marked BC there is also a contribution from their resonant annihilation via s-channel pseudoscalar and charged Higgs (A,H$^\pm$) states, like the funnel region of the CMSSM. There is an underabundance of dark matter relic density in the regions marked III, IV and V, and overabundance in the regions VI and VII. Table 5 lists the SUSY masses for a set of benchmark points in the region of gluino and stop masses ≤ 2 TeV, which can be probed at the LHC.

Table 5. SUSY masses for some benchmark points of Fig 10, corresponding to gluino and stop masses $\leq 2$ TeV [10].

| Parameter | 1 | 2 | 3 | 4 | 5 | 6 |
|---|---|---|---|---|---|---|
| $m_{1/2}$ | 848.94 | 818.46 | 874.26 | 833.33 | 732.93 | 881.86 |
| $m_0$ | 1663.05 | 2847.18 | 895.64 | 1102.62 | 2587.57 | 644.69 |
| $\tan\beta$ | 10.00 | 10.00 | 10.00 | 30.00 | 30.00 | 30.00 |
| $A_0$ | $-2000.00$ | $-2000.00$ | $-2000.00$ | $-2000.00$ | $-2000.00$ | $-2000.00$ |
| $(M_1, M_2, M_3)$ | 3781, 1364, 1779 | 3670, 1318, 1695 | 3880, 1402, 1842 | 3696, 1338, 1763 | 3272, 1181, 1534 | 3912, 1416, 1865 |
| $\mu$ | 1224.92 | 1096.58 | 1278.14 | 1180.55 | 962.14 | 1224.61 |
| $m_{\tilde{g}}$ | 1976.53 | 1969.82 | 1989.92 | 1911.12 | 1779.53 | 1993.02 |
| $m_{\tilde{\chi}_1^\pm}, m_{\tilde{\chi}_2^\pm}$ | 1215.42, 1420.40 | 1096.80, 1380.77 | 1264.42, 1453.96 | 1171.41, 1386.67 | 961.65, 1238.37 | 1216.89, 1459.56 |
| $m_{\tilde{\chi}_1^0}, m_{\tilde{\chi}_2^0}$ | 1214.18, 1235.18 | 1095.45, 1110.36 | 1263.25, 1286.26 | 1170.02, 1187.81 | 960.01, 974.02 | 1215.54, 1231.19 |
| $m_{\tilde{t}_1}, m_{\tilde{t}_2}$ | 1935.09, 2143.44 | 2413.91, 2711.43 | 1660.69, 2004.42 | 1577.30, 1957.53 | 2151.43, 2334.72 | 1512.69, 1995.50 |
| $m_{\tilde{b}_1}, m_{\tilde{b}_2}$ | 2002.13, 2467.81 | 2696.88, 3302.78 | 1699.20, 2111.78 | 1605.97, 1946.00 | 2298.94, 2754.09 | 1533.65, 1863.84 |
| $m_{\tilde{u}_1}, m_{\tilde{u}_2}$ | 2545.05, 3064.03 | 3362.44, 3751.19 | 2203.44, 2812.92 | 2213.15, 2771.82 | 3049.81, 3394.37 | 2137.80, 2770.35 |
| $m_{\tilde{e}_L}, m_{\tilde{e}_R}$ | 2510.23, 3534.60 | 3364.87, 4134.98 | 2140.52, 3338.13 | 2156.49, 3258.35 | 3047.09, 3731.57 | 2066.61, 3308.62 |
| $m_{\tilde{\tau}_1}, m_{\tilde{\tau}_2}$ | 2493.40, 3510.79 | 3346.80, 4105.56 | 2123.61, 3316.60 | 2001.68, 3055.40 | 2888.06, 3469.38 | 1907.25, 3112.53 |
| $m_A, m_{H^\pm}$ | 2757.87, 2757.70 | 3498.83, 3498.88 | 2462.24, 2462.44 | 2011.60, 2011.60 | 2612.90, 2612.90 | 1966.87, 1966.85 |
| $m_h$ | 122.42 | 122.01 | 122.90 | 123.42 | 122.45 | 123.77 |
| $\Omega_{\tilde{\chi}_1} h^2$ | 0.11 | 0.12 | 0.10 | 0.10 | 0.09 | 0.10 |
| $BF(b \to s\gamma)$ | $3.02 \times 10^{-4}$ | $3.04 \times 10^{-4}$ | $3.01 \times 10^{-4}$ | $2.78 \times 10^{-4}$ | $2.88 \times 10^{-4}$ | $2.78 \times 10^{-4}$ |
| $BF(B_s \to \mu^+\mu^-)$ | $3.53 \times 10^{-9}$ | $3.53 \times 10^{-9}$ | $3.54 \times 10^{-9}$ | $3.78 \times 10^{-9}$ | $3.79 \times 10^{-9}$ | $3.81 \times 10^{-9}$ |
| $\sigma_{pZ}^{SI}$ in pb | $1.41 \times 10^{-8}$ | $7.46 \times 10^{-9}$ | $1.65 \times 10^{-8}$ | $1.15 \times 10^{-8}$ | $6.54 \times 10^{-9}$ | $9.44 \times 10^{-9}$ |
| $\sigma_{\tilde{g}\tilde{g}}^{NLO}$ in fb | 1.09 | 1.26 | $9.73 \times 10^{-1}$ | 1.47 | 3.31 | $9.51 \times 10^{-1}$ |
| Dominant decay modes of $\tilde{g}$ in (%) (> 10% are shown) | $\tilde{g} \to \tilde{\chi}_1^0 t\bar{t}$ 12<br>$\to \tilde{\chi}_2^0 t\bar{t}$ 11<br>$\to \tilde{\chi}_1^- tb$ 22<br>$\to \tilde{\chi}_1^+ b\bar{t}$ 22 | $\tilde{g} \to \tilde{\chi}_1^0 t\bar{t}$ 20<br>$\to \tilde{\chi}_2^0 t\bar{t}$ 17<br>$\to \tilde{\chi}_1^- tb$ 23<br>$\to \tilde{\chi}_1^+ b\bar{t}$ 23 | $\tilde{g} \to \tilde{b}_1\bar{b}$ 29.5<br>$\to \tilde{b}_1^* b$ 29.5<br>$\to \tilde{t}_1\bar{t}$ 20.5<br>$\to \tilde{t}_1^* t$ 20.5 | $\tilde{g} \to \tilde{b}_1\bar{b}$ 29<br>$\to \tilde{b}_1^* b$ 29<br>$\to \tilde{t}_1\bar{t}$ 21<br>$\to \tilde{t}_1^* t$ 21 | $\tilde{g} \to \tilde{\chi}_1^0 t\bar{t}$ 17<br>$\to \tilde{\chi}_2^0 t\bar{t}$ 14<br>$\to \tilde{\chi}_1^- tb$ 23<br>$\to \tilde{\chi}_1^+ b\bar{t}$ 23 | $\tilde{g} \to \tilde{b}_1\bar{b}$ 26<br>$\to \tilde{b}_1^* b$ 26<br>$\to \tilde{t}_1\bar{t}$ 22<br>$\to \tilde{t}_1^* t$ 22 |
| Dominant decay modes of $\tilde{t}_1/\tilde{b}_1$ (in %) (> 10% are shown) | ... | ... | $\tilde{t}_1 \to t\tilde{\chi}_1^0$ 24<br>$\to t\tilde{\chi}_2^0$ 38<br>$\to b\tilde{\chi}_1^+$ 27<br>$\tilde{b}_1 \to t\tilde{\chi}_1^-$ 64<br>$\to t\tilde{\chi}_2^-$ 23 | $\tilde{t}_1 \to t\tilde{\chi}_1^0$ 24<br>$\to t\tilde{\chi}_2^0$ 36<br>$\to b\tilde{\chi}_1^+$ 30<br>$\tilde{b}_1 \to t\tilde{\chi}_1^-$ 61<br>$\to t\tilde{\chi}_2^-$ 14<br>$\to b\tilde{\chi}_1^0$ 10 | ... | $\tilde{t}_1 \to t\tilde{\chi}_1^0$ 23<br>$\to t\tilde{\chi}_2^0$ 42<br>$\to b\tilde{\chi}_1^+$ 33<br>$\tilde{b}_1 \to t\tilde{\chi}_1^-$ 74<br>$\to b\tilde{\chi}_1^0$ 14<br>$\to b\tilde{\chi}_2^0$ 10 |

One again sees an inverted squark mass hierarchy with relatively light stop, and higgsino dominance of the lighter chargino and neutrlino stated, marked by their degenerate masses. They lead to dominance of the gluino decay via (29), resulting in a similar SUSY signal as in the 75 model. Again with the predicted gluino pair production cross-sections of $\geq 1$ fb, one expects $\sim 100$ signal events at the high luminosity run of LHC. It should added here that there are proposals of increasing the energy of pp colliders to the 50-100 TeV range. It should be possible to probe the entire parameter space of the 75 and 200 models, shown in Figs 9 and 10, with a 50 TeV collider. The predicted direct dark matter detection cross-sections of these models are compatible with the present experimental limits; but most of their parameter spaces can be probed by the 1 Ton XENON experiment [10, 40].

## 7. Wino DM in Anomaly Mediated SUSY Breaking (AMSB) Model:

Let us finally consider the most popular NUGM model, based on anomaly mediated SUSY breaking [41]. Here the SUSY breaking in the hidden sector is communicated to the observable sector via the Super-Weyl anomaly contribution. This contribution is always present. Being a loop level contribution, however, this is suppressed relative to the tree level contribution of eq. (20). The AMSB model assumes that the SUSY breaking superfield does not belong to the singlet or (more generally) one of the representations occurring in the symmetric product of two adjoint representations (21) of the GUT group. Then the tree level contribution is forbidden by symmetry consideration, so that the AMSB contribution dominates. The resulting GUT scale gaugino and scalar masses along with the trilinear coupling parameters are given by the Callan-Symanzic $\beta$ and $\gamma$ functions times the gravitino mass $m_{3/2}$. i.e.

$$M_\lambda = \frac{\beta_g}{g} m_{3/2} \Rightarrow M_1 = \frac{33}{5} \frac{\alpha_1}{4\pi} m_{3/2}, M_2 = \frac{\alpha_2}{4\pi} m_{3/2}, M_3 = -3 \frac{\alpha_3}{4\pi} m_{3/2},$$

$$A_y = -\frac{\beta_y}{y} m_{3/2}, m_\phi^2 = -\frac{1}{4}\left(\frac{\partial \gamma}{\partial g} \beta_g + \frac{\partial \gamma}{\partial y} \beta_y\right) m_{3/2}^2 + m_0^2, \quad (30)$$

where g and y denote the gauge and Yukawa couplings. In the minimal version of this model (mAMSB) a common parameter $m_0^2$ is added to the GUT scale scalar particle mass squares to prevent the sfermion mass squares from turning negative (tachyonic) at the weak scale. All the weak scale superparticle masses are given by the RGE in terms of the three and half GUT scale parameters, $m_{3/2}$, $m_0$, $\tan\beta$ and sign of the higsino mass parameter $\mu$; while the magnitude of $\mu$ is again determined by the radiative electroweak symmetry breaking condition (5). One can easily see from the one-loop RGE (23) for the gaugino masses that at the weak scale the wino mass $M_2$ is the smallest of the three gaugino masses. We shall use the two-loop RGE code [11], in which case the weak scale gaugino masses are in the ratio

$$M_1 : M_2 : |M_3| \approx 2.8 : 1 : 7.1. \quad (31)$$

Fig 11 shows the $m_0$-$m_{3/2}$ parameter space of the mAMSB model for a representative value of $\tan\beta = 10$ [42]. The upper shaded region is disallowed because of no electroweak symmetry breaking, while the lower shaded region is disallowed because of sfermion LSP.

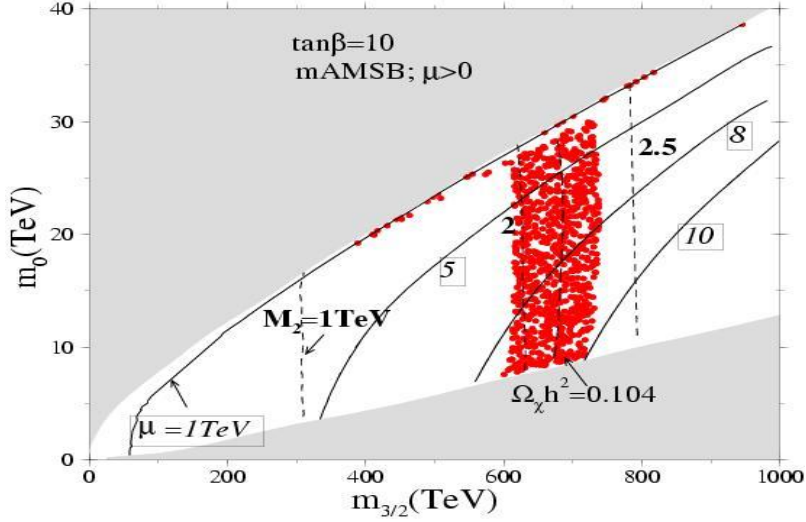

Fig 11. The $m_0 - m_{3/2}$ plot of the mAMSB model, with the cosmic dark matter relic density compatible region indicated by the red dots [42].

The cosmic dark matter relic density compatible region, computed with the the micrOMEGA code [32], is indicated by the red dots. The thin red band overlapping with the μ = 1 TeV contour corresponds to a 1 TeV higgsino LSP dark matter, annihilating via its isospin gauge coupling to the W boson as discussed earlier. The thick red band corresponds to a wino LSP dark matter, again annihilating via its isospin gauge coupling to W boson, i.e.

$$\tilde{W}^{\pm}\tilde{W}^0 \xrightarrow{W} \bar{f}f, \tilde{W}^0\tilde{W}^0 \xrightarrow{\tilde{W}^{\pm}} WW, \qquad (32)$$

proceeding via s-channel W boson and t-channel wino exchanges. Because the wino has a larger isospin (I = 1) gauge coupling compared to the higgsino (I = ½), the cosmic dark matter relic density is achieved for a larger wino LSP mass of

$$M_2 = 2.2 \pm 0.2 TeV. \qquad (33)$$

Therefore the SUSY mass spectrum of this model is beyond the reach of LHC. There is a viable anomalous single photon signature for wino pair production along with a photon at a future electron-positron collider like CLIC; but it will require a CM energy of at least 5 TeV [42].

There is an interesting indirect dark matter detection signal for wino dark matter, coming from their pair annihilation into W boson pair via wino exchange, i.e. the second process of eq. (32). The decay of these W bosons leads to a hard positron spectrum of the type observed by the PAMELA experiment [28]. More over one can boost the signal rate to the level observed by this experiment via Sommerfeld enhancement [43]. It arises from a distorted wave Born approximation calculation of the second process of eq. (32). Since the exchanged wino has a multi-TeV mass, the dark matter pair must come to contact (zero separation) for this Born annihilation process. On the way they experience the long range attractive force due to repeated exchange of the relatively light W boson, which enhances their contact wave function $\psi(0)$. The resulting enhancement of the Born annihilation cross-section by this $|\psi(0)|^2$ factor is the so called Sommerfeld enacement. This enhancement factor can be very large in the resonance region [43, 44].

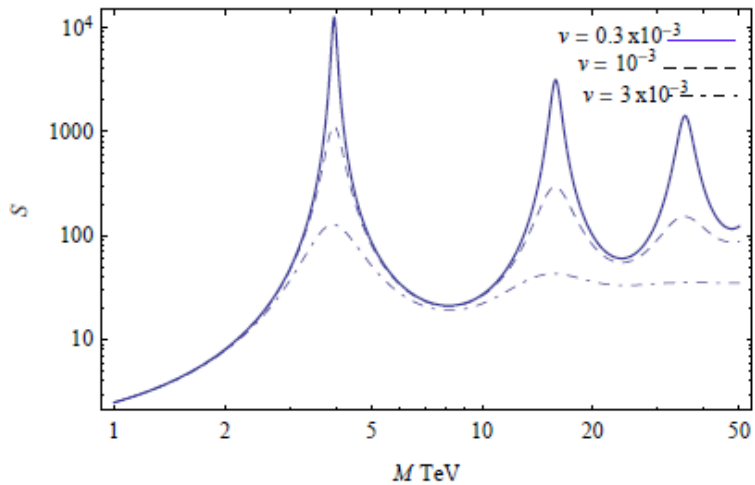

Fig 12. Sommerfeld enhancement from multiple W exchange as function of DM mass for different relative velocities [44].

Fig 12 shows this resonance to occur for a wino dark matter mass of 4 TeV; and the enhancement factor at this resonance can be as large as ten thousand for low DM velocities within the experimental range [44].

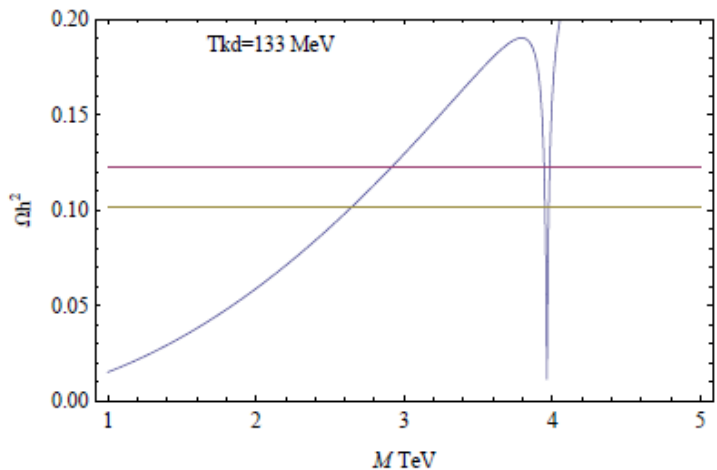

Fig 13. Effect of Sommerfeld enhancement on dark matter relic density [44].

Moreover, Fig 13 shows that incorporating the effect of Sommerfeld enhancement in the calculation of dark matter relic density [45], can bring it down to the cosmologically compatible range at the resonance mass of 4 TeV. Fig 14 shows that the resulting positron flux ratio is in good agreement with the PAMELA signal [28] both in shape and size.

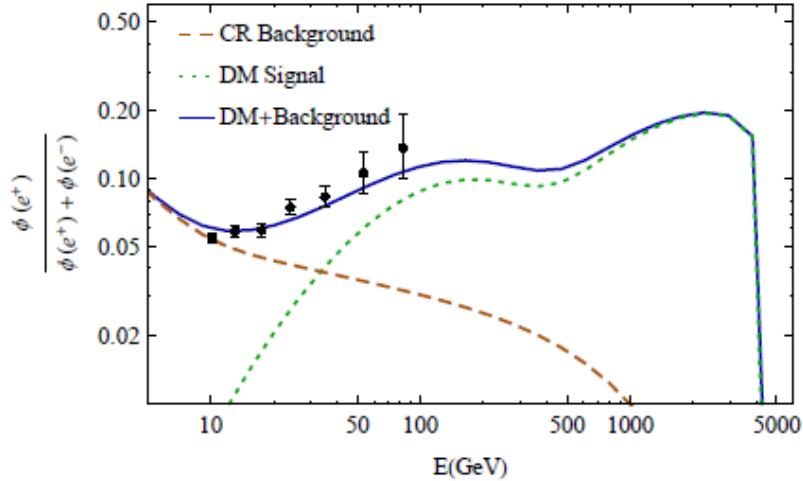

Fig 14. Positron flux ratio compared with PAMELA data as a function of energy [44].

Fig 15 compares the predicted antiproton/proton flux ratio with the corresponding PAMELA data [28]. Again there is reasonable agreement as the predicted peak is moved to higher energies.

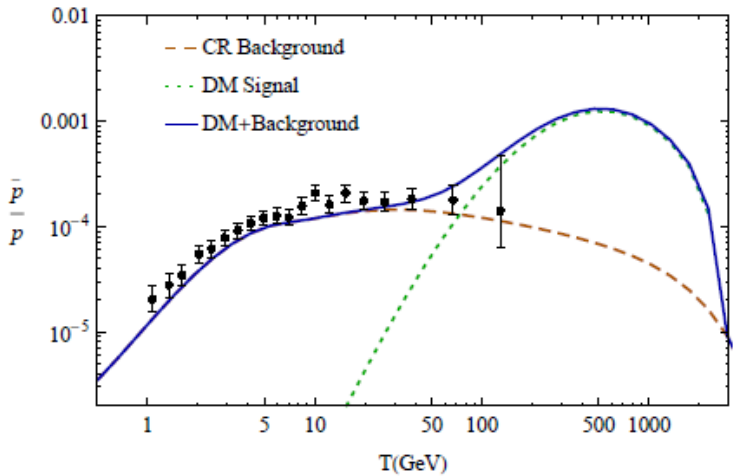

Fig 15. Antiproton/proton ratio compared with the PAMELA data as a function of energy [44].

In summary, assuming the fine-tuning involved in the resonant Sommerfeld enhancement, one can account for the shape and size of the PAMELA positron and antiproton fluxes in terms of a 4 TeV wino dark matter.

## 8. Summary:

In the MSSM the LSP dark matter can be an electroweak gaugino (bino, wino) or a higgsino. We have reviewed the phenomenology of this SUSY dark matter in several versions of MSSM with universal and nonuniversal gaugino masses at the GUT scale. We start with the universal case, called the CMSSM or mSUGRA model, which assumes SUSY breaking by a singlet superfield of the GUT group. Here the cosmologically compatible dark matter relic density is achieved only over four narrow regions of parameter space, each involving some fine-tuning of SUSY mass parameters, i.e. the stau coannihilation, resonant annihilation, focus point and higgsino LSP regions. Moreover, the first two are seriously challenged by the Higgs boson mass and the $B_S \to \mu^+\mu^-$ decay limit from the LHC, while the third one is challenged by the limits from the direct dark matter detection experiments. The fourth one is beyond the reach of LHC or direct dark matter detection experiments, and hence of little phenomenological interest. Then we discuss three types of simple and predictive nonuniversal gaugino mass (NUGM) models based on SU(5) GUT. The first type of models assume SUSY breaking by two superfields with a dominant singlet and a subdominant nonsinglet component, belonging to the 75-plet or 200-plet representations of SU(5). They can reconcile relatively light bino, wino, and sleptons with TeV scale gluino and squarks. The former ensures cosmologically compatible dark matter relic density through the pair annihilation of bino dark matter via slepton exchange (bulk annihilation process), while the latter ensures compatibility with the Higgs boson mass and other collider constraints. One of them, the 1+200 model, can even account for the muon anomalous magnetic moment. Both the 1+200 and 1+75 models can be probed at the LHC. The second type of NUGM models assume SUSY breaking by the two superfields, 1+75 or 1+200, with comparable singlet and nonsinglet components. The resulting dark matter is a mixed bino-higgsino state, which has sizable couplings to Z and Higgs bosons. So it can pair annihilate efficiently via these states, giving cosmologically compatible dark matter relic density over wide regions of parameter space. However, these models also predict large direct dark matter detection cross-sections via Higgs boson exchange, which is disfavoured by their current experimental limits. The third type of NUGM models assume SUSY breaking by a nonsinglet superfield, belonging to the SU(5) representations 75 or 200. They predict higgsino dark matter, which can pair annihilate efficiently via its Isospin (I = ½) gauge coupling to the W boson. One gets cosmologically compatible dark matter relic density for a higgsino mass of ~ 1 TeV, like the higgsino LSP region of the CMSSM; but the predicted gluino and squark masses are not so large as in CMSSM. At least a part of this mass range can be probed at the LHC. Finally we consider the NUGM model arising from anomaly mediated SUSY breaking. It predicts a wino dark matter, which can pair annihilate even more efficiently via its higher Isopsin (I = 1) gauge coupling to the W boson. One gets cosmologically compatible dark matter relic density for a wino mass of a little above 2 TeV, which puts it beyond the reach of LHC or direct dark matter detection experiments. But it has interesting predictions for some indirect dark matter experiments. In particular, it can account for the PAMELA events in shape and size if one assumes resonant Sommerfeld enhancement. There are many other NUGM models, which are beyond the scope of this review. Let us conclude by simply listing some the recent phenomenological studies of these models [46].

**Acknowledgement:** The works reported here were partially supported by a Senior Scientist Fellowship of the Indian National Science Academy.